\documentclass[doublecol]{epl2}
\usepackage{graphicx}

\title{{Theory of the} STM detection of Wigner molecules in spin incoherent {CNTs}}
\shorttitle{STM detection of Wigner molecules in spin incoherent CNTs} 

\author{N. Traverso Ziani\inst{1,2} \and F. Cavaliere\inst{1,2} \and M. Sassetti\inst{1,2}}
\shortauthor{N. Traverso Ziani \etal}

\institute{
  \inst{1} Dipartimento di Fisica, Universit\`a di Genova, Via Dodecaneso 33,
  16146, Genova, Italy.\\
  \inst{2} CNR-SPIN, Via Dodecaneso 33,
  16146, Genova, Italy.
}

\abstract{The linear conductance of a carbon nanotube quantum dot in the Wigner molecule regime, coupled to two scanning tunnel microscope tips is inspected. Considering the high temperature regime, the nanotube quantum dot is described by means of the spin-incoherent Luttinger liquid picture. The linear conductance exhibits spatial oscillations induced by the presence of the correlated, molecular electron state. A power-law scaling of the electron density and of the conductance as a function of the interaction parameter are found. They confirm {local} transport as a sensitive tool to investigate the Wigner molecule. The double-tip setup allows to explore different transport regimes with different shapes of the spatial modulation, all bringing information about the Wigner molecule.}
\begin{document}

\maketitle
\section{Introduction}
The quantum properties of interacting electrons are determined by the competition between kinetic energy and Coulomb repulsion. In three (3D) and two dimensions (2D), when Coulomb interactions are dominant a Wigner crystal emerges~\cite{wigner}, while quantum fluctuations in one-dimension (1D) prevent the formation of a fully-fledged Wigner crystal~\cite{giamarchi}. In finite-size systems, regardless of the dimensionality, a Wigner molecule emerges when the electronic correlation length exceeds the size of the sample~\cite{wigmol1,wigmol2}. While in symmetrical 3D and 2D systems Wigner molecules are detected via density-density correlation functions~\cite{wigmol1,wigmol2}, in 1D {systems} electron correlations produce strong spatial density fluctuations, which signal the occurrence of a molecular state~\cite{kramer,silva,xia,schulz,safi,sablikov}. Several efforts have been devoted to study numerically the occurrence of Wigner molecules in quantum dots, both in 2D~\cite{2Dnum2,Egger99,2Dnum3,2Dnum5,2Dnum6,2Dnum7,serra,2Dnum10,maxwf,wigspec} as well as in 1D~\cite{szafran,wire3,polini,shulenburger,secchi1,sgm2,astrak}.\\
From the experimental point of view, carbon nanotubes (CNT)~\cite{char} are very promising candidates to show Wigner correlations, due to the possibility to create 1D quantum dots either via external gates~\cite{bock,tans} or {by} exploiting geometrical defects~\cite{postma}. Experimental evidence of 1D Wigner molecules has been claimed in CNT quantum dots via transport experiments~\cite{desh} and in quantum wires via optical spectroscopy~\cite{hiraki} or momentum-resolved tunneling experiments~\cite{aus}. Recent theoretical studies~\cite{AFM,sgm2,mantelli} in the {\em low} temperature regime{~\cite{giamarchi,schulz,safi,giacomo}} $k_{B}T\ll D_{\sigma,\rho}$, where $D_{\sigma}$ ($D_{\rho}$) is the spin (charge) bandwidth of the system, have shown that an AFM tip capacitively coupled to the electron density of a 1D dot is a sensitive tool to detect signatures of Wigner molecules, {since the renormalization induced in the chemical potential and in the linear conductance follows the profile of the density}. At {\em higher} temperatures $D_{\sigma}\ll k_{B}T\ll D_{\rho}$, it has also been shown that an STM tip can be effective in detecting Wigner correlations via conductance oscillations, in close agreement with the ones of the electron density~\cite{secchi2}. {This issue is recovered in our model}. Several of these proposals have employed numerical techniques~\cite{secchi2}, extremely accurate but restricted to a low number of particles up to $N\approx 10$ and a limited range of system parameters. Analytical approaches, such as those based on the Luttinger liquid (LL) model~\cite{giamarchi}, are precious tools {permitting the consideration of higher} numbers of particles and to study with {greater} flexibility the interplay between confinement and Coulomb interactions, and the transition towards a Wigner molecule~\cite{sgm2,mantelli}.\\

\noindent In this Letter we evaluate the linear conductance of a CNT quantum dot tunnel-coupled to two STM tips. We will focus on the spin-incoherent regime at temperatures $D_{\sigma}\ll k_{B}T\ll D_{\rho}$, employing an appropriate LL model~\cite{fiete1,glazman1,glazman2} which replaces the more conventional {\em spin-coherent} LL model~\cite{giamarchi}, valid for $k_{B}T\ll D_{\sigma,\rho}$. The spin-incoherent model has already been employed to study transport properties, playing a crucial role to explain the reduction from $2e^2/h$ to $e^2/h$ ($e$ the electron charge) in strongly interacting quantum wires~\cite{075}. A crucial feature of this model is the ability to describe a strongly interacting system in which a Wigner molecule is {\em a priori} assumed to be the ground state. This simplifies the form of the {Hamiltonian}, allowing a powerful analytical approach to the study of a strongly correlated system. One can estimate $D_{\sigma,\rho}\approx N E_{\sigma,\rho}$ with $E_{\sigma}\ll E_{\rho}$ the spin ($\sigma$) and charge ($\rho$) excitation energies. For small $N$, thus, the spin-incoherent regime can be reached at moderate temperatures~\cite{075}.\\
Our main findings are the following:\\

\noindent ($i$) Even in the strongly interacting regime, the linear conductance exhibits a marked power-law scaling with the interaction strength. This scaling is dramatically different from that of the electron density and {permits the extraction of} the interaction parameter from a transport experiment;\\

\noindent ($ii$) The two-tips setup {permits the study of} a wide range of transport regimes, with different transparencies of the tunneling barriers. In all of them, signatures of the Wigner molecule are unambiguously detected. \\

\noindent The above findings confirm {local} transport as a sensitive tool to investigate Wigner molecules in a CNT.
\section{CNT Quantum dot in the spin-incoherent regime}
\label{sec:isolated}
The system under investigation is a strongly interacting CNT dot of length $L$. Electrons are considered to be in the Wigner molecule regime, which {permits their description as being} arranged on a linear chain and free to oscillate around their equilibrium positions with an antiferromagnetic nearest-neighbours interaction~\cite{glazman2}. To be more specific we review how this procedure is carried out for a system of spinful 1D electrons~\cite{glazman2}. The extension needed to deal with the CNT band structure will be introduced later.\\
The starting point is the standard {Hamiltonian} $H_F$ describing the low energy properties of the one dimensional quantum dot (from now on, $\hbar=1$)
\begin{eqnarray}
  H_F&=&\!\!\!\!\int_0^L {\mathrm d}x\ \psi^\dag_s(x)\epsilon(-i\partial_x)\psi_s(x) \label{eq:h}\\
  &+&\!\!\!\!\frac{1}{2}\int_0^L \mathrm{d}x\int_0^L \mathrm{d}y\ \psi^\dag_s(x)\psi^\dag_{s'}(y)V(x-y)\psi_{s'}(y)\psi_{s}(x)\, .\nonumber
\end{eqnarray}
Here, $\psi_s(x)$ is the electron operator with spin $s=\pm$, $\epsilon(k)$ is the dispersion relation, $V(x)$ is the repulsive interaction potential, and a sum over repeated indices is implied. When interaction is strong, electrons tend to be tightly confined on lattice sites and can be treated as spinless fermions, called {\em holons}, described by a Fermi operator $\Psi(x)$, and {Hamiltonian}
\begin{eqnarray}
  H_\rho&=&\int_0^L \mathrm{d}x\ \Psi^\dag(x)\epsilon(-i\partial_x)\Psi(x)\\
  &+&\frac{1}{2}\int_0^L \mathrm{d}x\int_0^L \mathrm{d}y\ \Psi^\dag(x)\Psi^\dag(y)V(x-y)\Psi(y)\Psi(x)\, .\nonumber
\end{eqnarray}
Despite the strong repulsive interaction, a weak residual antiferromagnetic interaction $H_\sigma$ among the spins of electrons is still present and described by
\begin{equation}
 H_\sigma=J\sum_{l=1}^{N-1} \textbf{S}_{l+1}\cdot \textbf{S}_{l}\, ,\label{eq:accasigma}
\end{equation}
where $N$ is the total number of electrons, $J$ is a (positive) exchange constant and $\textbf{S}_{j}$ is the spin operator of the $j$-th electron in the chain.\\
In the limit of strong interaction and for $D_{\sigma}\ll k_{B}T\ll D_{\rho}$, the {Hamiltonian} of the system simplifies and one has $H_{F}\approx H=H_{\rho}+H_{\sigma}$. The condition $D_{\sigma}\ll k_{B}T$ allows to perform the limit $J\to 0$ in every thermal average of system observables. The above {Hamiltonians} and the prescription on the thermal average constitute the {\em spin-incoherent} LL model~\cite{glazman2}.\\

\noindent The electron operator $\psi_s(x)$ can be expressed in terms of the holons as~\cite{glazman2}.
\begin{equation}
\psi_s^{\dagger}(x)=\Psi^{\dagger}(x)Z_{l(x),s},
\end{equation}
where $Z_{l,s}$, the operator that adds an electron with spin $s$ between the $l$-th and the $l+1$-th site of the chain and
\begin{equation}
l(x)=\int_0^{x}\mathrm{d}x'\ \Psi^\dag(x')\Psi(x')\, .\label{eq:l(x)}
\end{equation}
counts the number of holons in the domain $[0,x]$.
Note that the validity of this expression is limited to the case of strongly interacting electrons since one finds $\psi_s(x)\psi_{-s}(x)=0$, as expected for infinite repulsive interaction. Details can be found in ref.~\cite{glazman2}.\\
The charge sector can be further simplified: the bosonization identities for spinless fermions on a segment of length $L$, with open boundaries conditions~\cite{open}, allow to write the low energy sector of the {Hamiltonian} $H_\rho$ as
\begin{equation}
H_\rho=\frac{E_{\rho}}{2}\left(N-N^{(0)}-N_g\right)^2+\sum_{n_{q}>0}\varepsilon_{\rho}n_{q}d^\dag_{n_{q}}d_{n_{q}}, \label{eq:accarho}
\end{equation}
where $N^{(0)}$ is the total number of electrons in the reference state, $N_g$ is the charge induced by a gate contact capacitively coupled to it, $n_q$ are integers, $d_{n_{q}}$ are bosonic operators. The zero-mode and plasmon energies $E_\rho=2\pi v_{F}/Lg^2$ and $\epsilon_\rho=g E_{\rho}$ are written in terms of the Fermi velocity $v_F$ and of the LL interaction parameter $g$ with $0<g\leq 1$ for repulsive interactions. The interaction strength increases as $g$ gets smaller. The fermion field for the holons can be expanded as
\begin{equation}\Psi(x)=e^{ik_F^hx}\Psi_{+}(x)+e^{-ik_F^hx}\Psi_{-}(x)\, , \label{eq:optot}
\end{equation}
where $\Psi_{r}(x)$ are $2L$-periodic fermion fields representing right ($r=+$) and left ($r=-$) moving holons in the dot and $k_F^h=\pi N^{(0)}/L$ is their Fermi momentum. The open boundary conditions $\Psi(0)=\Psi(L)=0$ imply $\Psi_{r}(x)=-\Psi_{-r}(-x)$. The right movers operator admits a bosonic representation~\cite{open}
\begin{equation}
\Psi_{+}(x)=\frac{1}{\sqrt{2\pi\chi}}e^{-i\theta} \,e^{i\frac{\pi \Delta Nx}{L}}e^{i{\Phi_{\rho}(x)}}\,. \nonumber
\end{equation}
Here, $\chi=L/\pi N$ is the cutoff length, $\Delta N=N-N^{(0)}$, and $\theta$ satisfies $[\theta, \Delta N]=i$. The boson field $\Phi_{\rho}(x)$ is given by
\begin{equation}
\Phi_{\rho}(x)\!\!=\!\!\sum_{n_{q}>0}\frac{e^{-\chi q/2}}{\sqrt{g n_q}} \left\{\left[\cos{(qx)}-ig\sin{(qx)}\right]d^\dag_{n_{q}}+\mathrm{h.c.}\right\}\nonumber
\end{equation}
with $q=n_q\pi/L$. A complete basis for the eigenstates of the quantum dot is $|N,\left\{ o_q \right\},h_j\rangle,$ with $\left\{ o_q \right\}$ the occupation numbers of the collective holon modes, and $h_j$ the eigenstates of $H_{\sigma}$ {Hamiltonian}.\\

\noindent A generalization of the above results can be employed to describe a CNT, characterized by an extra quantum number $\alpha=\pm$ which labels the two non-equivalent $K$ points of the graphene honeycomb lattice. This quantum number can be formally associated to an isospin operator $\mathbf{\tau}$ corresponding to the angular momentum of electrons along the direction of the {circumference} of the CNT. Electrons in the CNT can thus be described as holons, located around lattice sites, with residual spin and isospin interactions. Also this theory can be bosonized. The holons {Hamiltonian} is still given by Eq.~(\ref{eq:accarho}), with interaction strength $g_{h}$. A connection can be made between this parameter and $g_{C}$, the interaction parameter of the {\em spin-coherent} LL theory for CNTs. One has $g_{h}={1}{/}{\sqrt{1+\frac{V_{0}}{4\pi v_{F}}}}$ {and} $\ g_{C}={1}{/}{\sqrt{1+\frac{4V_{0}}{\pi v_{F}}}}$ ,\label{eq:relation}
where $V_{0}$ is the long-wave Fourier transform of the Coulomb interaction, {so that} $g_{C}^{-2}=16g_{h}^{-2}-15$. Clearly, for $g_{h}\ll 1$ one has $g_C=g_h/4$. Typical values for the CNT interaction parameter are around $g_{C}\approx 0.2$~\cite{postma}. The electron operator per spin $s$ and in the valley $\alpha$ is now given by $\psi^{\dagger}_{s,\alpha}(x)=\Psi^{\dagger}(x)Z_{l(x),s,\alpha}$,
where $l(x)$ is still given by eq.~(\ref{eq:l(x)}), and $Z_{l,s,\alpha}$ are defined as the operators that add an electron with spin $s$ and valley index $\alpha$. The {Hamiltonian} that generalizes Eq.~(\ref{eq:accasigma}), whose expression will not be reported here, is composed by three terms describing all the possible residual exchange interactions involving the spin $\mathbf{S}$ and/or the isospin $\mathbf{\tau}$~\cite{nat}. All the exchange energies are supposed to be characterized by a single parameter $J$, as also confirmed by the spin-coherent theory for $H_\sigma$, described in terms of a three channel LL~\cite{desh}.\\
\noindent In order to confirm that the model that we introduced describes strongly interacting electrons in the Wigner molecule regime let us now consider the average electron density $\rho(x)=\sum_{s,\alpha}\langle\psi_{s,\alpha}^{\dagger}(x)\psi_{s,\alpha}(x)\rangle$,
where the brackets $\langle \cdot \rangle$ denote a thermal average.
Generalizing results for the 2-channels model~\cite{glazman2} one can show that $Z_{l(x),s,\alpha}Z_{l(x),s,\alpha}^{\dagger}=1/4$ which allows {us} to conclude that $\rho(x)\equiv\langle \Psi^\dag(x)\Psi(x)\rangle$.\\
\noindent In the following, we will be only interested in energy scales much smaller than the the energy $\varepsilon_{\rho}$, which allows {us} to evaluate the average in the zero temperature limit. Standard bosonization techniques allow to perform the calculation, and one finds
\begin{eqnarray}
\rho(x)&=&\frac{N}{L}\left\{1-\cos\left[\frac{2\pi Nx}{L}-2h(x)\right]d^{g_h}(x)\right\},\\
h(x)&=&\frac{1}{2}\tan^{-1}\left[\frac{\sin\left(\frac{2\pi x}{L}\right)}{e^{\pi\chi/L}-\cos\left(\frac{2\pi x}{L}\right)}\right],\\
d(x)&=&\left[\frac{\sinh\left(\frac{\pi\chi}{L}\right)}{\sqrt{\sinh^2\left(\frac{\pi\chi}{L}\right)+\sin^2\left(\frac{\pi x}{L}\right)}}\right].
\end{eqnarray}
\begin{figure}[htbp] \begin{center}
  \includegraphics[width=8cm,keepaspectratio]{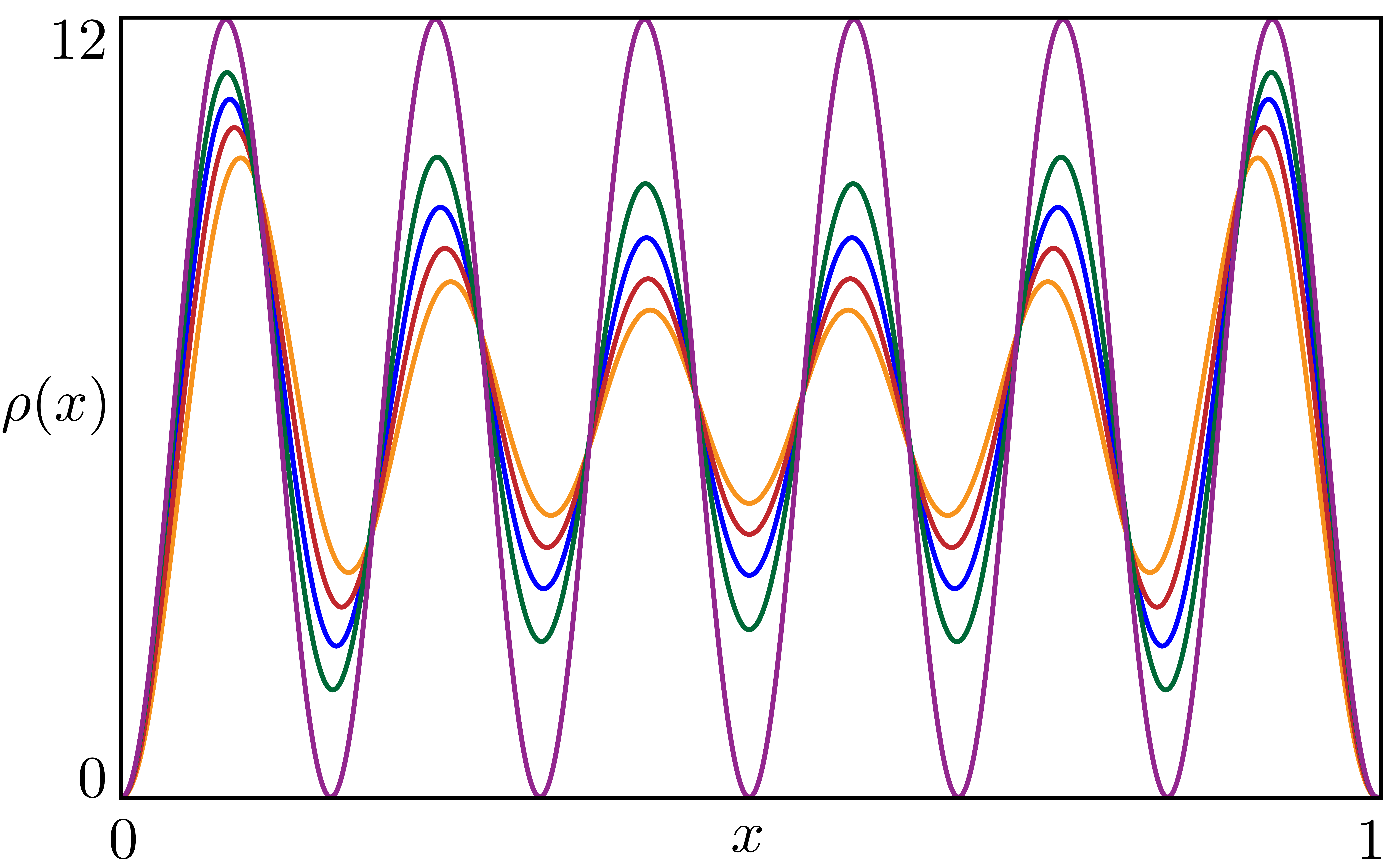}
  \caption{Electron density $\rho(x)$ (units $1/L$) as a function of $x$ (units $L$), for $N=6$ and different values of the CNT interaction parameter $g_C$: 1 (orange), 0.3 (red), 0.2 (blue), 0.1 (green), 0 (purple).}
  \label{fig:fig1}
  \end{center}
\end{figure}
The electron density $\rho(x)$ is shown in fig.~\ref{fig:fig1} for different values of $g_C$. As expected, it shows a distinct oscillating pattern with $N$ maxima {corresponding to the equilibrium positions of the electrons}, a clear signature of the presence of a Wigner molecule in the ground state of the system. {These maxima are, in the strong interaction limit, at the positions $m_j=jL/(N+1)+L/(2N)$, with $j=0,..,N-1$.} The case of noninteracting electrons $g_{C}=1$ is shown as {a} reference. As interactions increase, the density oscillations get more pronounced and in the extreme limit $g_{C}\to 0$, $\rho(x)$ exhibits fully developed and {saturated} maxima and minima where it attains respectively the value $2N/L$ and $0$. The scaling of $\rho(x)$ as a function of the interaction strength is dictated, for $g_C\ll 1$ by the function $d^{g_{h}}(x)\approx d^{4g_{C}}(x)$,
which suppresses the full oscillations of the density. In the range of interactions of physical interest, around $g_{C}\approx 0.2$, the oscillating pattern is only weakly affected by the above power-law scaling. As {will} be shown below, transport properties exhibit a much more dramatic dependence on the interaction parameter $g_{C}$.
\section{Transport properties}
\label{sec:trans}
We now turn to the transport properties of the CNT dot.
\begin{figure}[htbp]
\begin{center}
  \includegraphics[width=8cm,keepaspectratio]{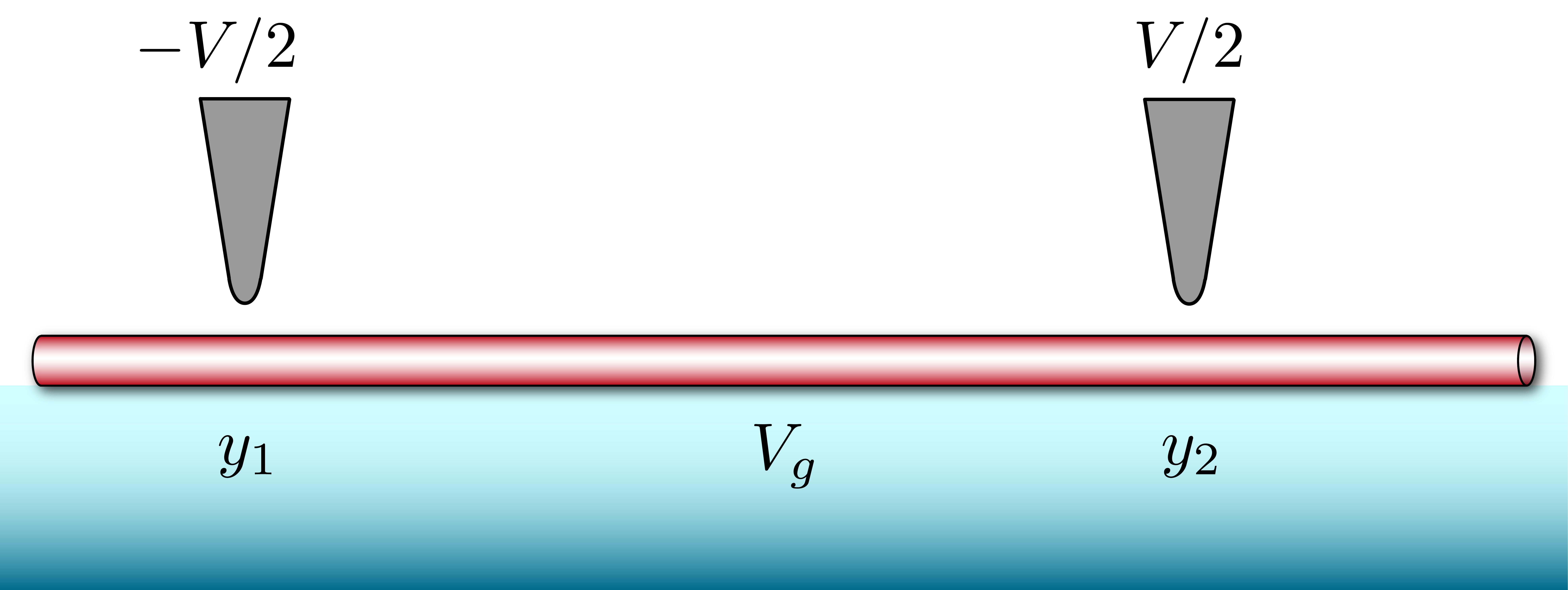}
  \caption{Schematic description of the set up: the CNT is coupled to two STM tips, at positions $y_1$ and $y_2$. Voltages are also indicated.}
  \label{fig:fig2}
  \end{center}
\end{figure}
Figure~\ref{fig:fig2} shows the setup that we will consider. It consists of a CNT dot deposed on a substrate acting as a back-gate, capacitively coupled to the dot. The back-gate is kept at a potential $V_{g}$ and induces a charge $eN_{g}=C_{g}V_{g}$ on the dot ($C_{g}$ is the gate-CNT capacitance), see Eq.~(\ref{eq:accarho}). The CNT is tunnel-coupled to two STM tips, modeled as semi-infinite non interacting, unpolarized, Fermi contacts with {Hamiltonian} $H_{tip}^{(i)}$ ($i=1,2$), placed above the CNT at a position $0\leq y_i\leq L$ along it. The tips are kept at a voltage $V_{i}=(-1)^{i}V/2$. The tunneling {Hamiltonians} read
\begin{equation}
H_{t}^{(i)}=\tau_i\sum_{s,\alpha} \psi^\dag_{s,\alpha}(y_i)\psi_{s,F}^i(0^+)+\mathrm{h.c.}\, ,\label{eq:tunham}
\end{equation}
where $\psi_{s,F}^i(z_i)$ are the Fermi field operators for the forward modes of the tip ($z_i$ is the coordinate along the tip with $z_i = 0$ at the vertex)~\cite{bercioux,noi2} with $\tau_i$ the transparencies of the tunneling barriers.\\
In this paper we will concentrate on linear transport in the sequential tunneling regime. We begin by evaluating the tunneling rates $\Gamma^{(i)}_{N\rightarrow N'}$ for the transition from a state of the CNT with $N$ electrons to a state with $N'$ occurring on the $i$-th barrier. In the sequential regime, one has $|N-N'|=1$ and Eq.~(\ref{eq:tunham}) can be treated to the lowest perturbative order. Furthermore the tip, the holonic excitations and the excitations of the spin and valley degrees of freedom in the CNT will be assumed in thermal equilibrium both in the initial and the final state. With these assumptions the tunneling rates can be evaluated as~\cite{AFM}
\begin{equation}
\Gamma^{(i)}_{N\rightarrow N+1}=|\tau_i|^2\int_{-\infty}^\infty \mathrm{d}t\ \langle Q_{s,\alpha}^{\dag}(t)
\rho(0)
Q_{s,\alpha}(0)\rangle_{\beta}\,,\label{eq:grule}
\end{equation}
where $Q_{s,\alpha}(t)=\psi_{s,\alpha}(y_i,t) \psi_{s,F}^{\dag}(0^+,t)$, $\langle O\rangle_{\beta}=\mathrm{Tr}\ \langle N+1|O|N+1\rangle$ with the trace performed over the degrees of freedom of the tips, the collective holon states and the eigenstates of $H_{\sigma}$. For definiteness, only the expression for a transition $N\to N+1$ has been quoted here. Crucially, due to the spin-incoherent regime, the tunneling rates {\em do not} depend on the spin $s$ or on the valley index $\alpha$ of the tunneling electron.\\
\noindent The time evolution in Eq.~(\ref{eq:grule}) is taken with respect to $H_{\sigma}+H_{tip}+H_{b}$ with $H_{tip}=H_{tip}^{(1)}+H_{tip}^{(2)}$ and $H_{b}$ the bosonic part of the holon {Hamiltonian} Eq.~(\ref{eq:accarho}). The density matrix of the initial state $\rho(0)$ is defined by
\begin{equation}
\rho(0)=\frac{e^{-\beta(H_\sigma+H_{tip}^{(i)}+H_b)}}{Z_{tip}Z_{\sigma}Z_b}|N\rangle\langle N|,
\end{equation}
where $Z_\nu$ ($\nu\in\{tip,b,\sigma\}$) are the partition functions of the {Hamiltonians} $H_\nu$.\\
The evaluation of the rate is straightforward in the spin-incoherent regime $J\rightarrow 0$, where the time evolution of the electron operator is given by~\cite{glazman2}
\begin{equation}
\psi_{s,\alpha}^{\dagger}(x,t)= \Psi^{\dagger}(x,t)Z_{l(x),s,\alpha}.
\end{equation}
In the linear regime, one finds
\begin{equation}
\Gamma^{(i)}_{N\rightarrow N+1}=\frac{\Gamma_{0}^{(i)}}{(N+1)^{g_{h}}}\varphi(y_{i})S(y_{i})f(\epsilon+eV_{i})\, , \label{eq:tunrate}
\end{equation}
where $\Gamma_{0}^{(i)}=\pi|\tau_i|^2\nu_{tip}^{(i)}(N+1)/2L$, $\nu_{tip}^{(i)}$ is the density of states in the $i$-th tip, $\varphi(y_{i})=d(y_i)^{-\frac{1}{2g_h}+\frac{g_h}{2}}\approx d(y_{i})^{-\frac{1}{8g_{C}}}$ for $g_{C}\ll 1$, and $S(y_{i})=\sin^{2}\left[\pi(N+1)y_{i}/L\right]$. Furthermore, $f(E)=\left[1+e^{\beta E}\right]^{-1}$ is the Fermi function with $\beta^{-1}=k_{B}T$ and $\epsilon=E_{\rho}\left(N-N^{(0)}-N_{g}+1/2\right)$. In the following we will assume $\nu_{tip}^{(1)}=\nu_{tip}^{(2)}$. The tunneling rates in Eq.~(\ref{eq:tunrate}) exhibit periodic spatial modulations given by $S(y_{i})$, enveloped by the function $\varphi(y_{i})$.\\
\noindent At the resonance between $N$ and $N+1$ electrons, occurring when $\epsilon=0$, the linear conductance $G=\lim_{V\to 0}\partial I/\partial V$ ($I$ the sequential current through the CNT) reads~\cite{master}
\begin{equation}
G=G_{0}\frac{1}{\Gamma_{0}^{(2)}}\frac{\Gamma^{(1)}_{N\rightarrow N+1}\Gamma^{(2)}_{N\rightarrow N+1}}{\Gamma^{(1)}_{N\rightarrow N+1}+\Gamma^{(2)}_{N\rightarrow N+1}}\ \ \ \ (\epsilon=V=0)\, ,\label{eq:lincond}
\end{equation}
where $G_{0}=2\beta e^{2}\Gamma_{0}^{(2)}$.
\section{Results}
\label{sec:results}
In the following we will assume the tip 1 fixed at a given position and study the linear conductance $G$ as a function of $y_{2}$, the position of the tip 2. The double-tip setup allows {us} to consider several different regimes, according to the ratio $A=\Gamma_{0}^{(1)}/\Gamma_{0}^{(2)}$. We begin by investigating the limit $A\gg 1$, where it is easy to show that
\begin{equation}
G\approx\frac{G_{0}}{(N+1)^{g_{h}}}\varphi(y_{2})\sin^{2}\left[\frac{\pi(N+1)y_{2}}{L}\right]\, ,\label{eq:Gas}
\end{equation}
for $\left|y_{1}-y_{0}^{(n)}\right|\gg L/\pi A(N+1)$ with $y_{0}^{(n)}=n{L}/(N+1)$ and $0\leq n\leq N+1$ an integer. Thus, the conductance in this regime is essentially independent of the position of the tip 1, except when the latter is at the border of the CNT or in the close proximity of one of the equilibrium positions of the electrons in the Wigner molecule with $N$ electrons, $y_{0}^{(1)}$,\ldots,$y_{0}^{(N)}$. Thus, the linear conductance exhibits the same spatial modulation pattern as the tunneling rate.
\begin{figure}[htbp]
\begin{center}
\includegraphics[width=8cm,keepaspectratio]{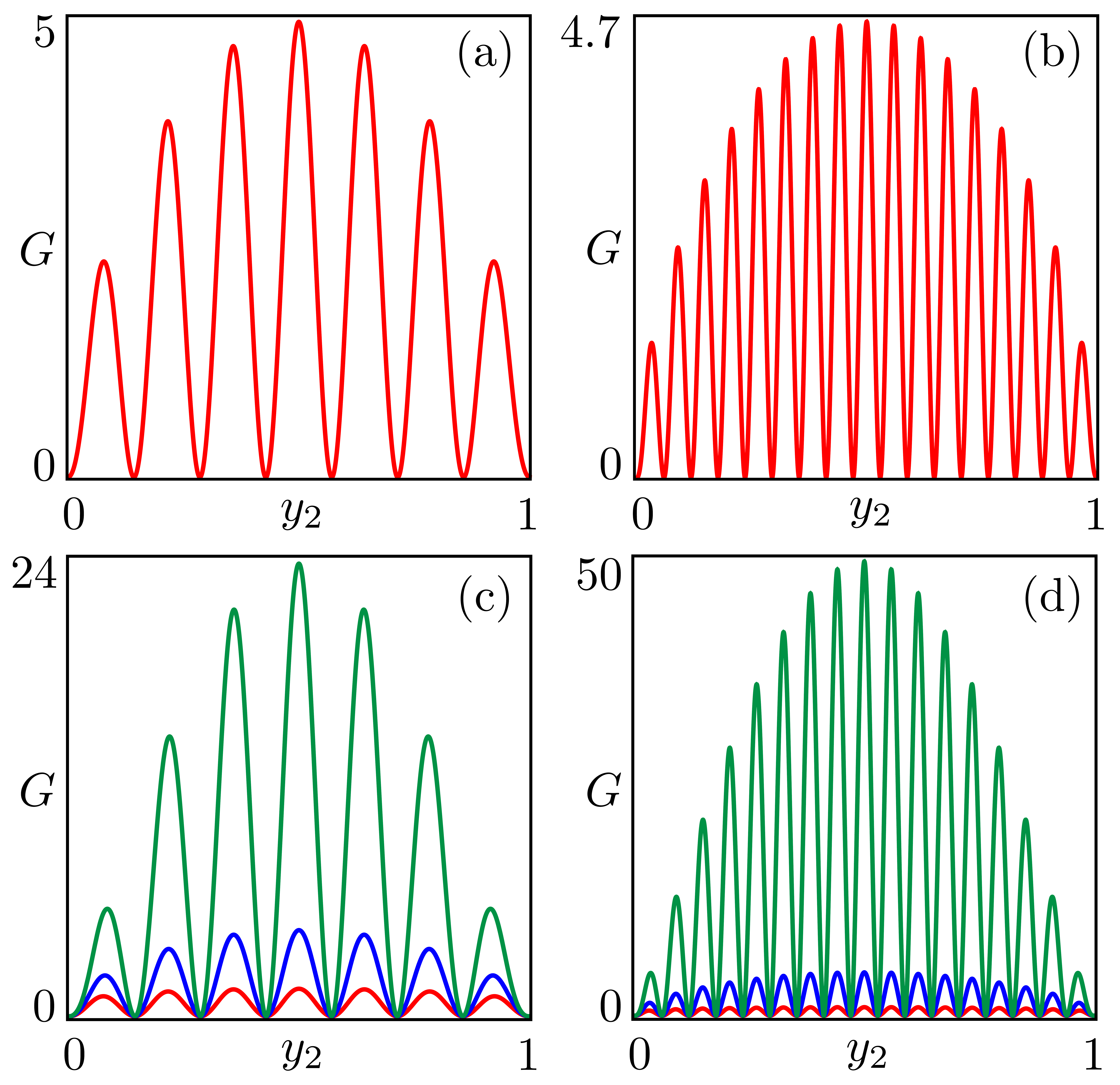}
\caption{Linear conductance $G
$ at the resonance between $N$ and $N+1$ charge states in the CNT (units $G_{0}$) as a function of the position of the tip 2 $y_{2}$ (units $L$). (a) Case of $N=6$ and $g_{C}=0.2$; (b) Case of $N=16$ and $g_{C}=0.2$; (c) Same as in (a) but for different values of $g_{C}$: 0.3 (red), 0.2 (blue) and 0.1 (green); (d) same as in (b) but for different values of $g_{C}$: 0.3 (red), 0.2 (blue), 0.1 (green). In all panels, $y_{1}=L/2$ and $A=20$.}
\label{fig:fig3}
\end{center}
\end{figure}
Figure~\ref{fig:fig3}(a) shows a plot of $G$ as a function of $y_{2}$. In {close} analogy to the behaviour of the electron density $\rho(x)$ {for $N+1$ electrons in the CNT}, strong modulations of the conductance occur, with $N+1$ maxima and $N$ nodes (barring the borders of the dot). These oscillations, stemming from the factor $\sin^{2}\left[\pi(N+1)y_{2}/L\right]$ in Eq.~(\ref{eq:Gas}) are due to the presence of Wigner molecules in the ground states for $N$ and $N+1$ electrons in the CNT. We therefore confirm numerical predictions~\cite{secchi2} which show how the linear conductance of a CNT probed by an STM tip in the high temperature regime displays evidence of the Wigner molecule. The power of the analytical method presented here, however, allows {us} to investigate higher numbers of particles, a regime not easily accessed by methods such as exact diagonalizations. As an example, Fig.~\ref{fig:fig3}(b) shows the conductance for the resonance between 16 and 17 electrons in the dot. The oscillating pattern is clearly present also for higher numbers of particles.\\
\noindent The oscillating patterns of the density are enveloped by the factor $\varphi(y_{2})$ which scales as $d(y_{2})^{-8g_{C}^{-1}}$. Thus, one can expect a marked dependence of $G$ on $g_{C}$: indeed, Fig.~\ref{fig:fig3}(c,d) show results for different values of the interaction strength. As $g_{C}$ decreases, the conductance oscillations grow without saturating, in contrast with the behaviour of $\rho(x)$. This important result, independent of the number of electrons in the dot, confirms that {local transport in the presence of STM tips} is a very sensitive tool to analyze electron correlations. It is to be understood here that even though $G$ increases as $g_{C}$ gets smaller, the amplitude of the tunnel barriers are always thought to be small enough, so as to ensure that $G\ll e^{2}$ and that the sequential tunneling approximation still holds.\\
\begin{figure}[htbp]
\begin{center}
\includegraphics[width=8cm,keepaspectratio]{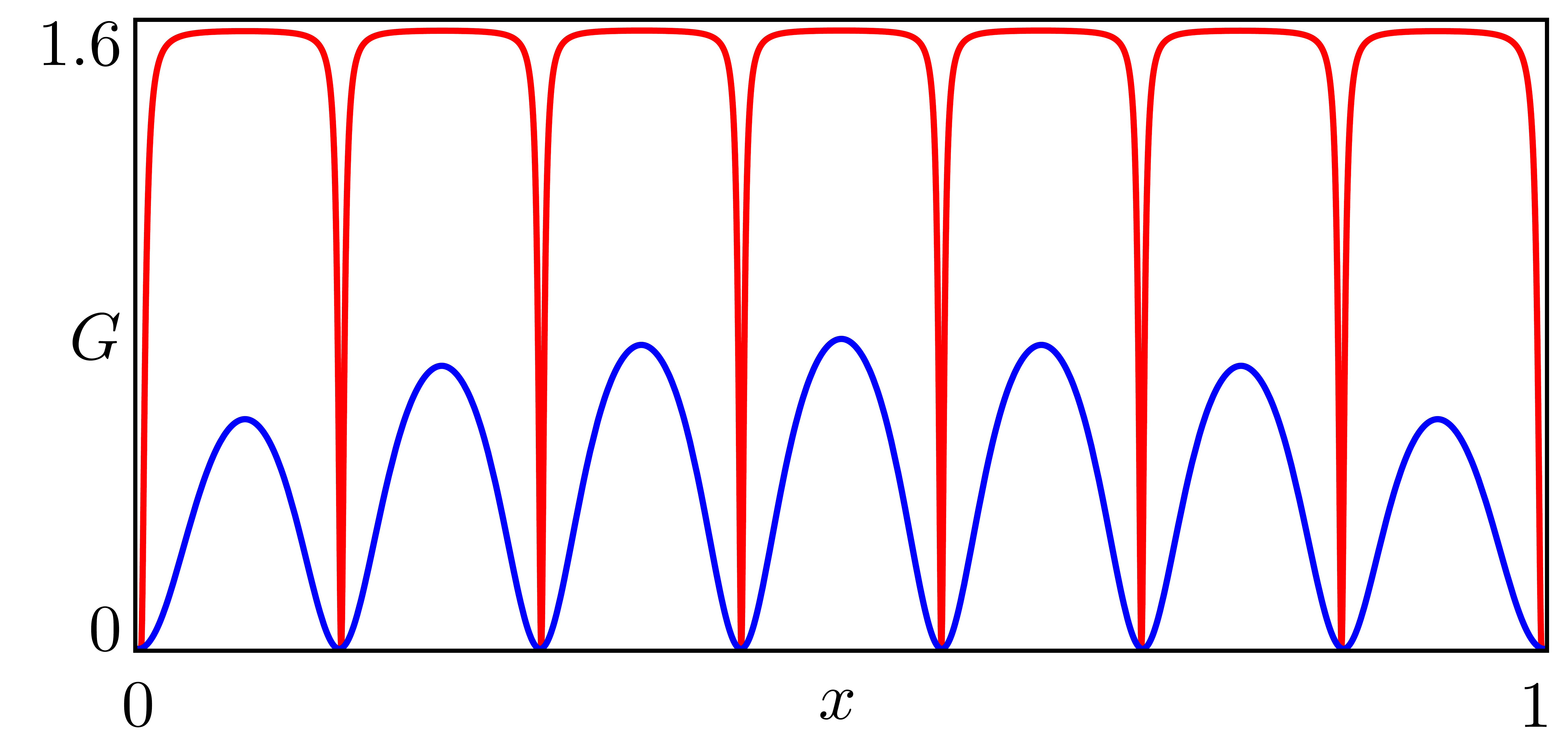}
\caption{Linear conductance $G$ at the resonance between states with 6 and 7 electrons in the CNT (units $AG_{0}$) as a function of the position of the tip 2 $y_{2}$ (units $L$) for different values of the asymmetry $A$: 1/20 (red) and 1 (blue). In all panels, $g_{C}=0.3$ and $y_{1}=L/2$.}
\label{fig:fig4}
\end{center}
\end{figure}
Let us now briefly address the other regimes of tips asymmetry. Figure~\ref{fig:fig4} shows the case $A=1$ (blue curve), whose qualitative behaviour is essentially identical to that of the asymptotic regime shown in fig.~\ref{fig:fig3}. More interesting is the case of $A\ll 1$, when the conductance can be approximated as
\begin{equation}
G\approx\frac{AG_{0}}{(N+1)^{g_{h}}}\varphi(y_{1})\sin^{2}\left[\frac{\pi(N+1)y_{1}}{L}\right]\, ,\label{eq:Gas2}
\end{equation}
for $\left|y_{2}-y_{0}^{(n)}\right|\gg A L/\pi (N+1)$. In this regime, thus, the conductance is essentially independent of $y_{2}$ except for narrow regions around the positions $y_{0}^{(n)}$ where the rate through the tip 2 is $\approx 0$. The red curve in fig.~\ref{fig:fig4} clearly shows this behaviour: the linear conductance is now constant almost everywhere along the CNT and $G$ exhibits sharp dips at the border of the CNT and at the position of the $N$ electrons in the Wigner molecule. Even in this regime, then, a strong signature of the presence of a Wigner molecule emerges in the conductance.\\
\noindent We close by observing that the power law scaling commented in the $A\gg 1$ regime holds in the whole range of barriers asymmetry and allows, in principle, to extract the interaction parameter $g_{C}$ with a transport experiment.
\section{Conclusions}
\label{sec:conclusions}
In this paper we have analytically investigated a CNT in the Wigner molecule regime at high temperature. The system has been described by means of the spin-incoherent LL. Both the electron density and the linear conductance of a sample probed by means of two STM tips exhibit spatial oscillations induced by the pinning of electrons along the equilibrium positions of the molecule. Our analytical approach has allowed to precisely identify the different power-law scalings of the density and conductance and to conclude that the spatial oscillations of the conductance are enhanced by the Coulomb interactions. The approach with two STM tips allows to address several different transport regimes where signatures due to the Wigner molecule are evident.
\acknowledgments
Financial support by the EU- FP7 via ITN-2008-234970 NANOCTM is gratefully acknowledged.


\begin{thebibliography}{99}
\bibitem{wigner}Wigner E., {\it Phys. Rev.}, {\bf 46} (1934) 1002.
\bibitem{giamarchi}Giamarchi T., \textit{Quantum Physics in One
  Dimension}, Oxford Science Publications (2004).
\bibitem{wigmol1} Reimann S. M. and Manninen M., {\it Rev. Mod. Phys.}, {\bf
  74} (2002) 1283.
\bibitem{wigmol2} Yannouleas C. and Landman U., {\it Rep. Prog. Phys.}, {\bf 70} (2007) 2067.
\bibitem{kramer}H\"ausler W. and Kramer B., {\it Phys. Rev. B}, \textbf{47} (1993)
  16353.
\bibitem{silva}Lima N. A. ,Silva M. F.,Oliveira,  L. N. and Capelle K., {\it Phys.
Rev. Lett.}, {\bf90} (2003) 146402.
\bibitem{xia}Xianlong G., {\it Phys. Rev. A}, \textbf{86} (2012)
  023616.
\bibitem{schulz}Schulz H. J., {\it Phys. Rev. Lett.}, {\bf 71} (1993) 1864.
\bibitem{safi}Safi I. and Schulz H. J., {\it Phys. Rev. B}, \textbf{59} (1999) 3040.
\bibitem{sablikov}Gindikin Y. and Sablikov V. A., {\it Phys. Rev. B}, \textbf{76} (2007) 045122.
\bibitem{2Dnum2} Hawrylak P. and Pfannkuche D., {\it Phys. Rev. Lett.}, {\bf 70} (1993) 485.
\bibitem{Egger99} Egger R., Hausler W., Mak C. H. and Grabert H., {\it Phys. Rev. Lett.},  {\bf 82} (1999) 3320.
\bibitem{2Dnum3} Tavernier M. B., Anisimovas E.,Peeters F. M.,
  Szafran B., Adamowski J. and Bednarek S., {\it Phys. Rev. B}, {\bf 68} (2003)
  205305.
\bibitem{2Dnum5} Rontani M., Cavazzoni C., Bellucci D. and
  Goldoni G., {\it J. Chem. Phys.}, {\bf 124} (2006) 124102.
\bibitem{2Dnum6} Harju A., Saarikoski H. and R\"as\"anen E.,
  {\it Phys. Rev. Lett.}, {\bf 96} (2006) 126805.
\bibitem{2Dnum7} Yannouleas C. and Landman U., {\it Phys. Rev. Lett.}, {\bf
  85} (2000) 1726.
\bibitem{serra} Puente A., Serra L. and Nazmitdinov R. G.,
  {\it Phys. Rev. B}, (2004) {\bf 69} 125315.
\bibitem{2Dnum10} Cavaliere F., De Giovannini  U., Sassetti M. and Kramer  B., {\it New J. Phys.}, {\bf 11} (2009) 123004.
\bibitem{wigspec} Kalliakos S., Rontani  M., Pellegrini V., Garcia C. P., Pinczuk A., Goldoni  G., Molinari E., Pfeiffer L. N. and
 West K. W., {\it Nat. Phys.}, {\bf 4} (2008) 467.
\bibitem{maxwf} Rontani M., Molinari  E., Maruccio  G., Janson  M.,
 Schramm A., Meyer C., Matsui T., Heyn C., Hansen W. and
 Wiesendanger R., {\it J. Appl. Phys.}, {\bf 101} (2007) 081714.
\bibitem{szafran} Szafran B., Peeters F. M., Bednarek S., Chwiej T.,
  and Adamowski J.,{\it Phys. Rev. B}, {\bf 70} (2004) 035401.
\bibitem{wire3} Mueller E. J. ,{\it Phys. Rev. B}, \textbf{72} (2005) 075322.
\bibitem{polini} Abedinpour S. H., Polini M., Xianlong G. and
 Tosi M. P., {\it Phys. Rev. A}, {\bf 75} (2007) 015602.
\bibitem{shulenburger} Shulenburger L., Casula M., Senatore G. and
 Martin R. M., {\it Phys. Rev. B}, {\bf 78} (2008) 165303.
\bibitem{secchi1} Secchi A. and Rontani M., {\it Phys. Rev. B}, \textbf{80} (2009)
  041404(R).
\bibitem{sgm2} Qian J., Halperin B. I. and Heller E. J., {\it Phys. Rev. B},
  \textbf{81} (2010) 125323.
\bibitem{astrak} Astrakharchik G. E. and Girardeau M. D., {\it Phys. Rev. B},
  {\bf 83} (2011) 153303.
\bibitem{char} Charlier J.-C., Blase X. and Roche S.,
  {\it Rev. Mod. Phys.}, \textbf{79} (2007) 677.
\bibitem{bock} Bockrath M., Cobden D. H., McEuen P. L.,
  Chopra N. G., Zettl A., Thess A. and Smalley R. E., {\it Science}, {\bf
    725} (1977) 1922.
\bibitem{tans} Tans S. J., Devoret M. H., Dai H., Thess A.,
  Smalley R. E., Geerligs L. J. and Dekker C., {\it Nature}, {\bf 386} (1997) 474.
\bibitem{postma} Postma H. W. C., Teepen T., Yao Z., Grifoni M. and
  Dekker C., {\it Science}, \textbf{293} (2001) 76.
\bibitem{desh} Deshpande V. V. and Bockrath M.,
  {\it Nat. Phys.}, \textbf{4} (2008) 314.

\bibitem{hiraki} Hiraki K. and Kanoda K.,
  {\it Phys. Rev. Lett.}, \textbf{80} (1998) 4737.
\bibitem{aus} Auslaender O. M., Steinberg H., Yacoby A., Tserkovnyak Y.,
   Halperin B. I., Baldwin K. W., Pfeiffer L. N. and West K. W.,
  {\it Science}, \textbf{308} (2005) 88.
\bibitem{AFM} Traverso Ziani N., Cavaliere F. and Sassetti M., {\it Phys. Rev. B}, \textbf{86} (2012) 125451.
\bibitem{mantelli}Mantelli D., Cavaliere F. and Sassetti M., {\it J. Phys.: Condens. Matter}, {\bf24} (2012) 432202.
\bibitem{giacomo}{Dolcetto G., Cavaliere F., Ferraro D. and Sassetti M., {\it Phys. Rev. B} \textbf{87} (2013) 085425.}
\bibitem{secchi2}Secchi A. and Rontani M., {\it Phys. Rev. B}, \textbf{85} (2012) 121410.
\bibitem{fiete1} Fiete G. A., Le Hur K. and Balents L.,{\it Phys. Rev. B}, \textbf{73} (2006) 165104.
\bibitem{glazman1} Matveev K. A., Furusaki A. and Glazman L. I.,
  {\it Phys. Rev. Lett.}, {\bf 98} (2007) 096403.
\bibitem{glazman2} Matveev K. A., Furusaki A. and Glazman L. I.,
  {\it Phys. Rev. B}, {\bf 76} (2007) 155440.
\bibitem{075} Matveev K. A., {\it Phys. Rev. Lett.}, {\bf 92} (2004) 106801.
\bibitem{open} Fabrizio M. and Gogolin A. O., {\it Phys. Rev. B}, \textbf{51} (1995) 17827.
\bibitem{nat} Deshpande V. V., Bockrath M., Glazman L. I. and Yacobi A., {\it Nature}, \textbf{464} (2010) 209.
\bibitem{bercioux} Bercioux D., Buchs G., Grabert H. and
  Gr\"oning O., {\it Phys. Rev. B}, \textbf{83} (2011) 165439.
\bibitem{noi2} Ziani N. T., Piovano G., Cavaliere F. and Sassetti M., {\it Physica Scripta}, {\bf T151} (2012) 014041.

\bibitem{master}Furusaki A., {\it Phys. Rev. B}, \textbf{57} (1998) 7141.

\end{thebibliography}
\end{document}